\documentclass[prl,twocolumn,showpacs,amsmath,amssymb]{revtex4}% Physical Review Letters
\usepackage{graphicx}% Include figure files
\usepackage{dcolumn}% Align table columns on decimal point
\usepackage{bm}% bold math
\begin{document}
\title{Analytic, Group-Theoretic
Wave Functions for Confined $N$-Body Quantum Systems}
\author{W.B. Laing$^1$}
\author{M. Dunn$^1$}
\author{J.G. Loeser$^2$}
\author{D.K. Watson$^1$}
\affiliation{$^1$University of Oklahoma, Department of Physics and
Astronomy, Norman, OK 73019 \\ $^2$Oregon State University,
Department of Chemistry, Corvallis, OR 97331}

\date{\today}

\begin{abstract}
%An abstract (double spaced) of no more than 600 characters, including
%spaces, which should be self-contained (no footnotes) for use in
%abstracting journals and databases.
%
Systems involving $N$-identical interacting particles under
quantum confinement appear in many areas of physics, including
chemical, condensed matter, and atomic physics. We discuss a
beyond-mean-field perturbation method that is applicable to weakly,
intermediate and strongly-interacting systems. Group theory is
used to derive an analytic beyond-mean-field correlated wave
function at zeroth order for a system under spherical confinement.
We derive the corresponding zeroth-order analytic density profile
and apply it to the example of a Bose-Einstein condensate.

\end{abstract}


\maketitle

\section{Introduction}
During the last two decades, novel $N$-body quantum systems have
been created using techniques to confine and manipulate atoms,
ions, and electrons. These systems, of both fundamental and
technological interest, include condensed atomic Bose gases, atoms
confined in optical lattices, quantum dots, and ultracold fermion
gases. Tuning the external fields of these environments
provides unique opportunities to study many-body effects over a
range of interaction strengths.

Mean-field treatments, such as the Hartree-Fock method in atomic
physics and the Gross-Pitaevskii method in condensed matter, do not include
correlation effects and fail to describe systems with tight confinement
or strong interaction. These systems, which have hundreds to
millions of particles, present serious challenges for existing
$N$-body methods, many of which were developed with small systems
in mind.

The methodology described below uses dimensional perturbation
theory (DPT)\cite{Copenhagen}, which has been previously applied to mostly 
small-$N$ systems in the form of
high-order, largely numeric calculations. In this letter, we use group theoretic techniques
to develop an {\em analytic} approach which fully exploits the
symmetry of the zeroth-order problem. This method avoids heavy numerical
computation, and $N$ enters into the theory as a parameter.
This crucial simplification allows results for any $N$ to
be obtained in a single calculation. This method also directly accounts for
each two-body interaction, rather than using an average interaction.
Even lowest-order results
include beyond-mean-field effects.
Thus, in
contrast to the low-density expansion methods pioneered by 
Lee, Huang and Yang in the 1950s\cite{LHY}, this method is
appropriate for the study of both weakly and strongly interacting
systems and the transition between them.
This general
formalism offers a systematic approach to the study of correlation
 in atomic and molecular, condensed-matter, chemical,
and nuclear systems.

Nearly all
past work using DPT has focused on  energies with little
attention given to wave functions. In this
paper we derive an {\em analytic},
correlated lowest-order $S$-wave wave function for $N$ identical
particles in a spherical confining potential. The lowest-order
wave function yields important information such as the nature
of excitations and expectation values of physical observables.
This result can be systematically improved by going to higher
order.

\section{Toolbox}
The tools used to describe large-$N$ correlated wave functions are
carefully chosen to maximize the use of symmetry and minimize the
dependence on numerical computation. We handle the massive number
of interactions for $N$ large ( $\sim N^2/2$ two-body
interactions) by bringing together three theoretical methods.

The first, DPT\cite{Copenhagen}, is chosen because its zeroth-order
equation, which is obtained for large $D$, yields a
maximally-symmetric configuration for $N$ identical particles. Higher
orders yield insight into fundamental motions as well as a framework
for successive approximations. 
The second method is the $FG$ method of Wilson, Decius, and
Cross\cite{wdc}. This seminal method
has long been used in quantum chemistry to study
vibrations of polyatomic molecules. It directly relates the
structure of the Schr\"odinger equation to the
coordinate set which describes the normal modes of the
system. The third method, the use of group
theoretic techniques\cite{wdc,hamermesh}, takes full advantage of
the symmetry at zeroth-order.

\subsection{Dimensional Perturbation Theory}
For $N$-body systems in large dimensions, the 
DPT wave function is localized about a symmetric structure
in which each particle is equidistant and equiangular
from every other particle. 
The Jacobian-weighted\cite{avery} DPT wave function is
harmonic and corresponds to oscillations about this structure.
Notwithstanding its relatively simple
form, the large-dimension, zeroth-order wave function
includes beyond-mean-field effects.

The Schr\"{o}dinger equation for the zeroth-order Jacobian-weighted wave
function has the form\cite{annals}
\begin{equation}\label{Gham} \left(-\frac{1}{2}
{\mathbf{\bm{\partial}_{\bar{y}'}}}^{T} {\mathbf G} \,
{\mathbf{\bm{\partial}_{\bar{y}'}}} + \frac{1}{2}
\bar{\mathbf{y}}^{\prime T} {\mathbf F} \, {{\bar{\mathbf{y}}'}} +
v_o \right) \, \Phi_0({\mathbf{\bar{y}'}}) = \overline{E}_0 \,
\Phi_0({\mathbf{\bar{y}'}}),
\end{equation}
where ${\bar{\mathbf{y}}'}$ is the displacement coordinate vector
formed from dimensionally-scaled internal displacement coordinates
$\bar{r}'_i$ ($1 \leq i \leq N$) and $\overline{\gamma}'_{ij}$ ($1
\leq i < j \leq N$). Coordinates $\bar{r}'_i$ and
$\overline{\gamma}'_{ij}$ are related to the $D$-dimensional
scalar radii $r_i$ of the $N$ particles from the center of the
confining potential and the cosines $\gamma_{ij}$ of the
$N(N-\nolinebreak 1)/2$ angles between the radial vectors by
\begin{equation} \label{eq:taylor}
r_i = \kappa(D) \, \left(\bar{r}_{\infty}+\delta^{1/2}\bar{r}'_i
\right), \hspace{1.5ex} \mbox{and} \hspace{1.5ex} \gamma_{ij} =
\overline{\gamma}_{\infty}+\delta^{1/2}\overline{\gamma}'_{ij} \,.
\end{equation}
The expansion parameter is $\delta=1/D$, and $\kappa(D)$ is quadratic
in $D$ with the particular form chosen to simplify the resulting
equations\cite{scalings}. The quantities
$\bar{r}_{\infty}$ and $\overline{\gamma}_{\infty}$ define the
symmetric large-dimension structure, which depends on the nature and strength of the
interparticle interaction and confining potential.
%In Eq.~(\ref{Gham})
The matrices ${\bf G}$ and ${\bf F}$  are constants derived from the Hamiltonian.

\subsection{The ${\bf F} \, {\bf G}$ Method}

The FG method\cite{wdc} is used to obtain 
normal-mode coordinates and frequencies from the eigenvalue problem
in Eq.~(\ref{Gham}). The
$b^{\rm th}$ normal-mode coordinate ${q'}_{b}$ may be written as

\begin{equation} \label{eq:qyt}
\renewcommand{\arraystretch}{1.5} \begin{array}{rl}
[{\mathbf q'}]_b = {\mathbf{b}}^T {\bar{\mathbf{y}}'} \,, &
\mbox{where} \hspace{1.5ex} {\bf F} \, {\bf G} \, {\mathbf{b}} =
\lambda_b \,
{\mathbf{b}} \,, \\
& {\mathbf{b}}^T {\bf G} \, {\mathbf{b}} = 1\,, \hspace{1.5ex}
\lambda_b = \bar{\omega}_b^2,
\end{array}
\end{equation}
and $\bar{\omega}_b$ is the normal-mode frequency. Equation~(\ref{eq:qyt}) 
still represents a formidable eigenvalue problem
unless $N$ is quite small, since there are $P = N(N+1)/2$ normal
coordinates and up to $P$ distinct frequencies.

\subsection{Group Theory: the $S_N$ Symmetry}
The full $S$-wave Hamiltonian is invariant under particle
interchange. This fact defines an $S_N$ symmetry under which the system is
invariant. As defined in Eq.~(\ref{eq:taylor}), the large-dimension
structure is a completely symmetric configuration so
Eq.~(\ref{Gham}) is also invariant under the group $S_N$. This
$S_N$ symmetry\cite{hamermesh} brings about a remarkable
reduction from $P$ possible distinct frequencies to five actual
distinct frequencies and greatly simplifies
the determination of the normal coordinates through the use of symmetry
coordinates.\cite{wdc}

As a prelude to the above, we note that the $S_N$ invariance of
Eq.~(\ref{Gham}) means that the ${\bf F}$, ${\bf G}$ and ${\bf
F}{\bf G}$ matrices of Eq.~(\ref{eq:qyt}) are invariant under
$S_N$, which implies that the eigenvectors ${\mathbf{b}}$ and normal
modes transform under irreducible representations (irreps.) of
$S_N$. Using the theory of group characters\cite{hamermesh}, the coordinates
$\bar{r}'_i$ are reduced to one 1-dimensional and one
$(N-\nolinebreak 1)$-dimensional irrep., labelled $[N]$ and
$[N-\nolinebreak 1,1]\,$ respectively\cite{annals}. The
$\overline{\gamma}'_{ij}$ are reducible to one 1-dimensional, one
$(N-\nolinebreak 1)$-dimensional, and one $N(N-\nolinebreak
3)/2$-dimensional irrep., labelled $[N]$, $[N-\nolinebreak
1,1]\,$, and
 $[N-\nolinebreak 2,2]\,$ respectively. Since the normal modes transform under
irreps.\ of $S_N$ and are composed of linear combinations of
elements of vectors $\bar{\mathbf{r}}'$ and
$\overline{\bm{\gamma}}'$\,, there will be two $1$-dimensional,
two $(N-\nolinebreak 1)$-dimensional, and one entirely angular
$N(N-\nolinebreak 3)/2$-dimensional irreps.\ labelled by the
partitions $[N]$, $[N-\nolinebreak 1,1]$ and  $[N-\nolinebreak
2,2]$ respectively. All normal modes that transform together under
the same irrep.\ have the same frequency, so rather than $P$
distinct frequencies there are only {\em five}!

\section{Normal Modes -- The Program}
We determine the normal coordinates and distinct frequencies in a
three-step process\cite{annals}:

First, we define sets of
primitive irreducible coordinates that have the simplest possible functional
form subject to the requirement that they transform under
particular non-orthogonal irreps.\ of $S_N$.
%For the $\bar{\mathbf{r}}'$ sector
We define two sets of linear
combinations of elements of the $\bar{\mathbf{r}}'$ vector which
transform under non-orthogonal $[N]$ and $[N-\nolinebreak 1,1]$
irreps.\ of $S_N$\,. We then derive two sets of linear
combinations of elements of the $\overline{\bm{\gamma}}'$ vector
which transform under exactly these same two irreps.\ of $S_N$.
Finally we define a set of linear combinations of elements of
$\overline{\bm{\gamma}}'$ which transform under a particular
non-orthogonal $[N-\nolinebreak 2,2]$ irrep.\ of $S_N$.

Second, we use linear
combinations within each set of primitive irreducible
coordinates to determine symmetry coordinates that are defined to
transform under {\em orthogonal} irreps.\ of $S_N$\,. Care is
taken to ensure that this transformation to the symmetry
coordinates preserves the identity of equivalent representations
in the $\bar{\mathbf{r}}'$ and $\overline{\bm{\gamma}}'$ sectors.
We choose one of the symmetry coordinates to be a single primitive
irreducible coordinate, the simplest functional form possible
that transforms irreducibly under $S_N$.
The next symmetry coordinate is chosen to be composed of two
primitive irreducible coordinates, and so on. Thus the complexity
of the symmetry coordinates is minimized, building up slowly as
symmetry coordinates are added.

Third, the ${\bf FG}$
matrix, which was originally expressed in the $\bar{\mathbf{r}}'$ and
$\overline{\bm{\gamma}}'$ basis, is now expressed in symmetry
coordinates. This results in a stunning simplification. The $P \times
P$ eigenvalue equation of Eq.~(\ref{eq:qyt}) is reduced to one $2
\times 2$ eigenvalue equation for the $[N]$ sector,
$N-\nolinebreak 1$ identical $2 \times 2$ eigenvalue equations for
the $[N-\nolinebreak 1,1]$ sector, and $N(N-\nolinebreak 3)/2$
identical $1 \times 1$ eigenvalue equations for the
$[N-\nolinebreak 2,2]$ sector. For the $[N]$ and $[N-\nolinebreak
1,1]$ sectors, the $2 \times 2$ structure allows for mixing of the
$\bar{\mathbf{r}}'$ and $\overline{\bm{\gamma}}'$ symmetry
coordinates in the normal coordinates (see Eq.~(\ref{eq:qS})
below). The $1 \times 1$ structure of the equations in the
$[N-\nolinebreak 2,2]$ sector reflects the absence of
$\bar{\mathbf{r}}'$ symmetry coordinates in this sector, i.e.\ the
$[N-\nolinebreak 2,2]$ normal modes are entirely angular.

%
%~~~~~~LOOK OUT! MANUAL PAGE BREAK!!!~~~~~~~~~~~~~~~~~
\pagebreak

\subsection{The Symmetry Coordinates.} Using
steps one and two, we derive the
symmetry coordinates
\begin{widetext}
%\onecolumngrid
\begin{equation} \label{eq:S}
\renewcommand{\arraystretch}{1.5}
\begin{array}{@{}r@{\hspace{0.5ex}}c@{}l@{}}
{\mathbf{S}}_{r'}^{[N]} & = & {\displaystyle \frac{1}{\sqrt{N}} \,
\sum_{k=1}^N \overline{r}'_k \,,} \hspace{2em}
{\mathbf{S}}_{\gamma'}^{[N]} = {\displaystyle
\sqrt{\frac{2}{N(N-1)}} \,\,\, \sum_{l=2}^N \sum_{k =1}^{l-1}
\overline{\gamma}'_{kl} \,,}  \hspace{2em}
\protect[{\mathbf{S}}_{r'}^{[N-1,1]}\protect]_i = {\displaystyle
\frac{1}{\sqrt{i(i+1)}} \left( \sum_{k=1}^i \overline{r}'_k - i
\overline{r}'_{i+1} \right)\,, %\mbox{\hspace{2ex} where
%\hspace{2ex}} 1 \leq i \leq N-1 \,,
} \\
%\protect[{\mathbf{S}}_{r'}^{[N-1,1]}\protect]_i & = &
%{\displaystyle \frac{1}{\sqrt{i(i+1)}} \left( \sum_{k=1}^i
%\overline{r}'_k - i \overline{r}'_{i+1} \right)\,,
%\mbox{\hspace{2ex} where \hspace{2ex}} 1 \leq i \leq N-1 \,,
%\,,} \\
\protect[{\mathbf{S}}_{\gamma'}^{[N-1,1]}\protect]_i & = &
{\displaystyle \frac{1}{\sqrt{i(i+1)(N-2)}} \, \left( \left[
\sum_{l = 2}^i \, \sum_{k=1}^{l-1} \hspace{-1ex}
\overline{\gamma}'_{kl} + \sum_{k = 1}^i \, \sum_{l=k+1}^{N}
\hspace{-1ex} \overline{\gamma}'_{kl} \right] - i \left[
\sum_{k=1}^i \overline{\gamma}'_{k,\,i+1} + \sum_{l=i+2}^N
\hspace{-0.5ex}
\overline{\gamma}'_{i+1,\,l} \right] \right) \,,} \\
\multicolumn{3}{l}{\hspace{2ex} \mbox{where} \hspace{2ex} 1 \leq i
\leq N-1 \,, \hspace{2ex} \mbox{and} } \\
\protect[{\mathbf{S}}_{\gamma'}^{[N-2,2]}\protect]_{ij} & = &
{\displaystyle \frac{1}{\sqrt{i(i+1)(j-3)(j-2)}} \, \left(
\vphantom{\sum_{k=1}^{[j'-1, i]_{min}} \hspace{-2ex}
\overline{\gamma}'_{kj'}} \right. }
\begin{array}[t]{@{}c@{}}
\displaystyle{ \hspace{-5ex} \sum_{j'=2}^{j-1} \sum_{k=1}^{[j'-1,
i]_{min}} \hspace{-2ex} \overline{\gamma}'_{kj'} +
\sum_{k=1}^{i-1} \sum_{j'=k+1}^i
\overline{\gamma}'_{kj'} - (j-3) \sum_{k=1}^i \overline{\gamma}'_{kj} - } \\
\left. \displaystyle{ - i \, \left( \sum_{k=1}^{i}
\overline{\gamma}'_{k,(i+1)} + \sum_{j'=i+2}^{j-1}
\overline{\gamma}'_{(i+1),j'} \right) + i (j-3)
\overline{\gamma}'_{(i+1),j} } \right)  \,,
\end{array}
\\
\multicolumn{3}{l}{\hspace{2ex} \mbox{where} \hspace{2ex} 1 \leq i
\leq j-3 \hspace{2ex} \mbox{and} \hspace{2ex} i+3 \leq j \leq N
\,. }
\end{array}
\renewcommand{\arraystretch}{1}
\end{equation}
%\twocolumngrid
\end{widetext}
%
%where $1 \leq i \leq j-3$ and $i+3 \leq j \leq N$\,.

\subsection{The Normal Coordinates.}
In the third step, Eq.~(\ref{eq:qyt}) is expressed in the
symmetry coordinate basis of Eq.~(\ref{eq:S}) and reduces to three
eigensystem equations ( $\alpha = [N], [N-1], [N-2,2]$ ):
\begin{equation} \label{eq:sceig}
\bm{\sigma}_{\alpha}^{FG} {\mathsf{c}}^{\alpha} = \lambda_\alpha
{\mathsf{c}}^{\alpha}, \hspace{2ex} \mbox{and} \hspace{2ex}
[{\mathsf{c}}^{\alpha}]^T \bm{\sigma}_{\alpha}^G
{\mathsf{c}}^{\alpha} = 1 \,.
\end{equation}
The reduced ${\bf F} \, {\bf G}$ matrices $\bm{\sigma}_{[N]}^{FG}$
and $\bm{\sigma}_{[N-\nolinebreak 1,1]}^{FG}$ are $2 \times 2$
matrices, while $\bm{\sigma}_{[N-\nolinebreak 2,2]}^{FG}$ is a $1
\times 1$ matrix. The same is true for the diagonal reduced ${\bf
G}$ matrices, $\bm{\sigma}_{\alpha}^G$\,. The elements of
$\bm{\sigma}_{\alpha}^{FG}$ and $\bm{\sigma}_{\alpha}^G$  are
known analytic functions\cite{annals}.

There are five solutions to Eq.~(\ref{eq:sceig}) denoted 
${\bf 0}^\pm = \{\lambda^\pm_{[N]},\,{\mathsf{c}}_\pm^{[N]}\}$\,,
${\bf 1}^\pm = \{\lambda^\pm_{[N-\nolinebreak 1,1]},
\,{\mathsf{c}}_\pm^{[N-\nolinebreak 1,1]}\}$ and ${\bf 2} =
\{\lambda_{[N-\nolinebreak 2,2]},\,{\mathsf{c}}^{[N-\nolinebreak
2,2]}\}$\,. The two-element ${\mathsf{c}}^{\alpha}_\pm$ vectors
for the $\alpha = [N]$ and $[N-\nolinebreak 1,1]$ sectors
determine the angular-radial mixing of the symmetry coordinates in
a normal coordinate of a particular $\alpha$\,. Hence
\begin{equation} \label{eq:qS}
[{\mathbf{q}'}]_b =
[{\mathsf{c}}^{\alpha}_\pm]_{\bar{\mathbf{r}}'} \,
[{\mathbf{S}}_{r'}^{\alpha}]_\xi \, + \,
[{\mathsf{c}}^{\alpha}_\pm]_{\overline{\bm{\gamma}}'} \,
[{\mathbf{S}}_{\gamma'}^{\alpha}]_\xi \,.
\end{equation}
The normal coordinate label $b$ is replaced by the labels
$\alpha$, $\xi$ and $\pm$ on the rhs of
Eq.~(\ref{eq:qS}). For the $[N-\nolinebreak 2,2]$ sector, the
symmetry coordinates are also the normal coordinates up to a
normalization constant, $[{\mathsf{c}}^{[N-\nolinebreak
2,2]}]_{\overline{\bm{\gamma}}'}$\,.

\subsection{The Wave Function.} \label{subsec:wavef}
The wave function in Eq.~(\ref{Gham}) is the product of
$P$ harmonic-oscillator wave
functions:
\begin{equation} %\label{eq:}
\Phi_0({\mathbf{\bar{y}'}}) = \prod_{ \mu = \{
\mathbf{0}^\pm,\hspace{0.5ex} \mathbf{1}^\pm,\hspace{0.5ex}
\mathbf{2} \}
}
 \,\, \prod_{\xi=1}^{d_{\mu}} \,\, \phi_{n_{\mu_\xi}}
\!\! \left( \sqrt{\bar{\omega}_\mu} \,\,\, [{\mathbf{q}'}^\mu]_\xi
\right) \,,
\end{equation}
where $\phi_{n_{\mu_\xi}} \!\! \left( \sqrt{\bar{\omega}_\mu}
\,\,\, [{\mathbf{q}'}^\mu]_\xi \right) $ is a one-dimensional
harmonic-oscillator wave function of frequency $\bar{\omega}_\mu$,
and $n_{\mu_\xi}$ is the oscillator quantum number, $0 \leq
n_{\mu_\xi} < \infty$, which counts the number of quanta in each
normal mode. The quantity $\mu$ labels the manifold of normal
modes with the same frequency $\bar{\omega}_\mu$ while $d_{\mu} =
1$\,, $N-\nolinebreak 1$ or $N(N-\nolinebreak 3)/2$ for $\mu =
{\bf 0}^\pm$\,, ${\bf 1}^\pm$ or ${\bf 2}$ respectively.

\section{Unmasking the Wave Function: The Density Profile}
In the case of macroscopic quantum-confined systems, such as a BEC,
the wave function is made manifest in the experimentally
accessible density profile.

The large-dimension Jacobian-weighted ground-state density profile,
$N_0(r)$\,, is an analytic function:
\pagebreak
\[
\renewcommand{\arraystretch}{1.5}
\begin{array}[b]{@{}l@{}}
{\displaystyle 4\pi \, N_0(r) = 4\pi \, r^2 \rho_0(r) } \\
{\displaystyle =  \sum_{i=1}^N \int_{-\infty}^\infty \cdots
\int_{-\infty}^\infty \delta(r-r_i) \,
[\Phi_0({\mathbf{\bar{y}'}})]^2 \hspace{-2ex} \prod_{\mu= {\bf
0}^\pm, {\bf 1}^\pm, {\bf 2}} \, \prod_{\xi=1}^{d_{\mu}}
d[{\mathbf{q}'}^\mu]_\xi } \\
{\displaystyle = N \sqrt{\frac{R}{\pi}} \, \exp{[-R \left(
r-\kappa(D)\bar{r}_{\infty} \right)^2 ], } \, } \hfill (9)
\end{array}
\renewcommand{\arraystretch}{1}
\]
\addtocounter{equation}{2}

\noindent where
$R$ depends on $N$, on the frequencies $\bar{\omega}_{{\bf 0}^\pm}$
and $\bar{\omega}_{{\bf 1}^\pm}$, and on the angular-radial
mixing coefficients
$[{\mathsf{c}}^{[N]}_\pm]_{\bar{\mathbf{r}}'}$\,,
$[{\mathsf{c}}^{[N]}_\pm]_{\overline{\bm{\gamma}}'}$\,,
$[{\mathsf{c}}^{[N-\nolinebreak 1,1]}_\pm]_{\bar{\mathbf{r}}'}$
and $[{\mathsf{c}}^{[N-\nolinebreak
1,1]}_\pm]_{\overline{\bm{\gamma}}'}$\,.

\subsection{An Example: the BEC Density Profile}
\begin{figure}
\includegraphics[scale=1.0]{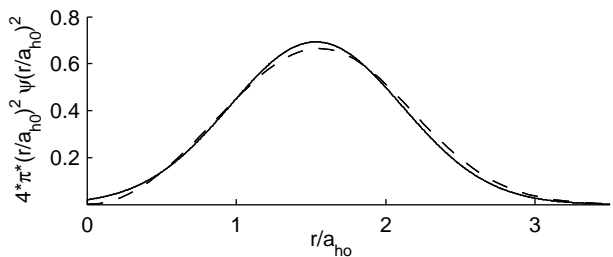}
\caption{Number density per atom versus radial distance of a spherically confined
BEC of 100 $^{87}$Rb atoms with $a=1\,000$ a.u.
and $\omega_{ho} = 2\pi \times 77.87$ Hz. The solid line is the
analytic 
DPT density and the dashed line is 
the MGP density.} \label{fig:large_a1}
\end{figure}

We perform a proof-of-concept test on
our zeroth-order density profile by considering a $T=0$K
BEC in an isotropic, harmonic trap.
We take the interatomic potential to be a hard sphere of radius
$a=0.0433a_{ho}$ (ten times the natural scattering length of
$^{87}$Rb)\, and $N = 100$ (a ten-fold increase in $N$ beyond existing
benchmark diffusion Monte Carlo (DMC)
density results\cite{dmc}). At  this $N$ and $a$
the modified Gross-Pitaevskii method (MGP)\cite{MGP}
is still valid for comparison.
To implement the perturbation theory with $\kappa(D) = D^2 a_{ho}$,
we allow this hard-sphere potential to soften away from $D=3$ by defining
the interatomic potential

%\begin{widetext}
\begin{equation} V_{\mathtt{int}}(r_{ij})  =\frac{\bar{V_0}}{(1-3/D)} \left[
1-\tanh\left[ \frac{\bar{c}_{o}\left(r_{ij}-\sqrt{2} \bar{a}\right)}{\sqrt{2} (1-3/D)}
 \right] \right],
\end{equation}
%\end{widetext}
where $\bar{a}=D^2a$, and the parameters $\bar{V_0}$ and $\bar{c_0}$
specify the height and width. We exploit this
freedom in the shape and position of the interatomic potential at
large $D$ to ensure that the lowest-orders energy and wave
function, $\Phi_0({\mathbf{\bar{y}'}})$, are as accurate as
possible. The parameters $\bar{V_0}$ and $\bar{c_0}$
    are determined by a least-squares fit to 
benchmark DMC energies ($N = 5 - 100$) and
densities ($N = 3$ and $10$)\cite{dmc}.
In Figure~\ref{fig:large_a1} our analytic density profile compares well
to the MGP result.

The analytic DPT density profile of Eq.~(9) is a Gaussian, symmetric about
$\kappa(D)\bar{r}_{\infty}$.
For $a$ or $N$ 
sufficiently large, the physical density profile develops an asymmetry
which motivates future work on the
next-order DPT wavefunction.

\section{Summary and Conclusions} \label{sec:SumConc}
This letter discusses an analytic approach to the study of
quantum-confined $N$-body systems.
Unlike
mean-field methods, this approach directly accounts for each
interparticle interaction, even at lowest-order, and  so is
applicable to high-density, strongly-interacting systems.  By taking advantage
of three powerful methods, DPT, the FG method and the group theory of the
$S_N$ group, we avoid heavy numerical computation and offer a systematic
approach to correlation that is not limited to small-$N$ systems.
While most
prior work with DPT has focused on energies, this letter derives
the lowest-order, analytic, correlated wave function for a
spherically-confined, $N$-particle system and the
corresponding analytic density
profile.

\section{Acknowledgments}

We acknowledge continued support from
the Army Research Office and the Office of
Naval Research.
We thank Doerte Blume for DMC results.


\begin{thebibliography}{99}
\bibitem{Copenhagen} D.R.\ Herschbach, J.\ Avery, and O.\
Goskinski, Eds., \textit{Dimensional Scaling in Chemical Physics}.
Kluwer Academic, Dordrecht, 1993.
\bibitem{LHY} K.\ Huang and C.N.\ Yang, Phys.\ Rev.\ {\bf 105},
767 (1957);
T.D.\ Lee and C.N.\ Yang, ibid. 1119;
%T.D.\ Lee and C.N.\ Yang, Phys.\ Rev.\ {\bf 105}, 1119 (1957):
T.D.\ Lee, K.\ Huang and C.N.\ Yang, Phys.\ Rev.\ {\bf
106}, 1135 (1957).
\bibitem{wdc} E.B.\ Wilson, Jr., J.C.\ Decius, P.C.\ Cross,
\textit{Molecular Vibrations: The Theory of Infrared and Raman
Vibrational Spectra}. McGraw- Hill, New York, 1955. (See especially
Appendix XII, p. 347)
\bibitem{hamermesh} M.\ Hamermesh, {\it Group theory and its
Application to Physical Problems}. Addison-Wesley, Reading, MA,
1962.
\bibitem{avery} J.\ Avery, D.Z.\ Goodson, D.R.\ Herschbach,
Theor.\ Chim.\ Acta \textbf{81}, 1 (1991).
\bibitem{annals} M.\ Dunn, D.K.\ Watson and J.G.\ Loeser, Ann.\
Phys.\ (NY), to be submitted.
\bibitem{scalings} Choices of $\kappa(D)$ for the $N$-electron
atom, the $N$-electron quantum dot and the BEC may be found in
B.A.\ McKinney, M.\ Dunn, D.K.\ Watson, and J.G.\ Loeser, Ann.\
Phys.\ (NY) \textbf{310}, 56 (2004).
\bibitem{dmc} D. Blume and C.H. Greene,
  {\sl Phys. Rev. A \bf 63}, 063601(2001).
Private communication, Doerte Blume.
\bibitem{MGP} E.\ Braaten and A.\ Nieto, Phys.\ Rev.\ B {\bf 56},
14745 (1997); E.\ Timmermans, P.\ Tommasini and K.\ Huang, Phys.\
Rev.\ A {\bf 55}, 3645 (1997).
\end{thebibliography}
\end{document}